\documentclass{article}

    \PassOptionsToPackage{square, numbers, sort&compress}{natbib}

 \usepackage[dblblindworkshop, final]{neurips_2025}
\workshoptitle{Machine Learning and the Physical Sciences}

\usepackage[utf8]{inputenc} %
\usepackage[T1]{fontenc}    %
\usepackage{hyperref}       %
\usepackage{url}            %
\usepackage{booktabs}       %
\usepackage{amsfonts}       %
\usepackage{nicefrac}       %
\usepackage{microtype}      %
\usepackage{xcolor}         %
\usepackage{graphicx}
\usepackage{subcaption}
\usepackage{float}

\newcommand{\eg}{\emph{e.g.}}
\newcommand{\etc}{\emph{etc.}}
\newcommand{\vs}{\emph{vs.}}
\newcommand{\dcpy}{\texttt{DistClassiPy}}
\newcommand{\dimmad}{DiMMAD}

\title{In Search of the Unknown Unknowns: A Multi-Metric Distance Ensemble for Out of Distribution Anomaly Detection in Astronomical Surveys} %

\author{%
  Siddharth Chaini\thanks{\href{https://sidchaini.github.io/}{sidchaini.github.io}} \\
  University of Delaware \\
  \texttt{\href{mailto:chaini@udel.edu}{chaini@udel.edu}} \\
  \AND
  Federica B. Bianco \\
  University of Delaware \\
  Rubin Observatory \\
  \And
  Ashish Mahabal \\
  Caltech \\
}

\begin{document}

\maketitle

\begin{abstract}
Distance-based methods involve the computation of distance values between features and are a well-established paradigm in machine learning. In anomaly detection, anomalies are identified by their large distance from normal data points. However, the performance of these methods often hinges on a single, user-selected distance metric (\eg, Euclidean), which may not be optimal for the complex, high-dimensional feature spaces common in astronomy. Here, we introduce a novel anomaly detection method, Distance Multi-Metric Anomaly Detection (\dimmad), which uses an ensemble of distance metrics to find novelties.  

Using multiple distance metrics is effectively equivalent to using different geometries in the feature space. By using a robust ensemble of diverse distance metrics, we overcome the metric-selection problem, creating an anomaly score that is not reliant on any single definition of distance. We demonstrate this multi-metric approach as a tool for simple, interpretable scientific discovery on astronomical time series -- (1) with simulated data for the upcoming Vera C. Rubin Observatory Legacy Survey of Space and Time, and (2) real data from the Zwicky Transient Facility. 

We find that \dimmad{}
excels at out-of-distribution anomaly detection -- anomalies in the data that might be new classes -- and beats other state-of-the-art methods in the goal of maximizing the diversity of new classes discovered. For rare in-distribution anomaly detection, \dimmad{} performs similarly to other methods, but may allow for improved interpretability. All our code is open source: \dimmad{} is implemented within \dcpy: \href{https://github.com/sidchaini/distclassipy/}{https://github.com/sidchaini/distclassipy/}, while all code to reproduce the results of this paper is available here: \href{https://github.com/sidchaini/dimmad/}{https://github.com/sidchaini/dimmad/}.

\end{abstract}

\section{Introduction}
The imminent Vera C. Rubin Observatory's Legacy Survey of Space and Time (LSST) is poised to revolutionize time-domain astronomy, generating time series alerts capturing changes in brightness for millions of objects every night \cite{ivezicLSSTScienceDrivers2019}. While LSST's unique capability -- its high resolution, large depth of view, and high cadence -- will enable unprecedented studies of known phenomena, its greatest promise lies in its potential to discover entirely new astrophysical objects, often referred to as the `unknown unknowns' or `true novelties' \cite{li2021preparing}. Historically, these discoveries have often been serendipitous, and have drastically improved our physical understanding of the Universe (\eg{} pulsars \cite{hewish_observation_1968}, `Oumuamua \cite{meech_brief_2017}, \etc). However, the sheer scale and volume of LSST data requires us to `plan for serendipity', where  machine learning will be crucial in enabling the discovery of unusual astrophysical phenomena \cite[\eg ][]{lochner_astronomaly_2021, malanchev_anomaly_2021,martinez-galarza_method_2021, villar_deep-learning_2021,muthukrishna_real-time_2022, volnova_snad_2024,perez-carrasco_multi-class_2023, gagliano_physics-informed_2023, aleo_anomaly_2024, gupta_classifier-based_2025, kornilov_coniferest_2025, romao_anomaly_2025, pruzhinskaya_what_2025}.

The task of finding the novel phenomena can be framed as an anomaly detection (AD) problem, and can be seen through two different lens: (1) Out of Distribution (OOD) detection -- identifying objects from a very different class than a training set of known classes (\eg, finding supernovae when trained on variable stars), and (2) Rare In-Distribution detection -- identifying extreme or rare examples which are more like a different subclass rather than a new class in itself (\eg, finding superluminous supernovae when trained on common supernovae). While both are important to astrophysics research, OOD detection offers the best chance at finding the `unknown unknowns' \cite{yang_generalized_2024} to enable the discovery of fundamentally new physics.

The search for these anomalies in time-domain astronomy is typically a two-stage process. The raw, irregularly-sampled time series data (light curves) must be encoded into a fixed-length feature vector. This can be accomplished through feature extraction \cite[\eg ][]{sanchez-saezAlertClassificationALeRCE2021, malanchev_light-curve_2021} or learned via deep learning methods such as representation learning \cite[\eg ][]{villar_deep-learning_2021, parker_astroclip_2024}. Then, an anomaly detection algorithm is applied to this feature space to identify outliers (some common methods are described later in \autoref{sec:exp}).

A significant challenge, however, is that standard AD algorithms can struggle to prioritize OOD objects \cite[\eg ][]{aleo_anomaly_2024}. 
This can obscure true novelties by saturating the list of top anomaly candidates with statistically rare and interesting, but astrophysically known objects. To address this, we propose a novel, semi-supervised AD method for identifying anomalies from a given feature representation: Distance Multi-Metric Anomaly Detection (\dimmad{}).

\section{Methods}
Our anomaly detection method \dimmad{} builds on \dcpy, a distance-based classifier that uses multiple distance metrics \cite{chaini_light_2024}. The authors successfully applied it to the classification of variable stars by extracting variability features from light curves. Their classification is based on the `distance' of the test object features to the centroids of training object features. With the use of the right distance metric for a classification problem, their classification performance matched the state-of-the-art, while being significantly faster and more interpretable. However, they showed that the performance hinges on choosing the right distance metric for a given classification problem, as the best choice depends on the topology of the feature space. Here, we extend this method to anomaly detection, and use an ensemble of distance metrics to overcome the metric selection problem altogether.

\subsection{Distance metrics}
\begin{figure}[htb!] %
    \centering
    \includegraphics[height=0.5\textwidth]{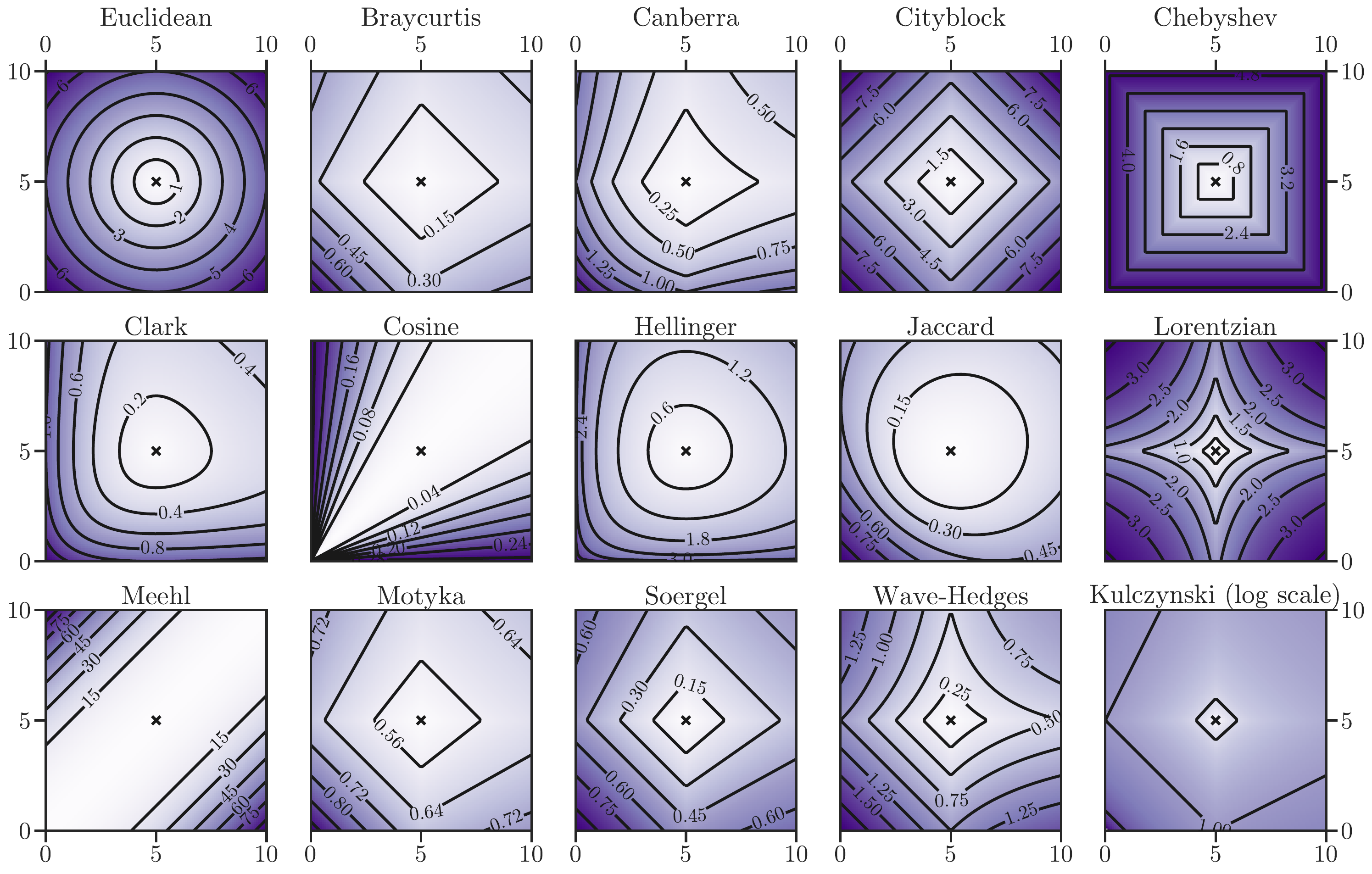}
    \caption{A visualization of 15 (of total 16) distance metrics used in our anomaly detector, \dimmad{}. In each panel, the intensity of the color denotes the distance value from a central point $(5,5)$ for that metric in a 2-dimensional feature space. Contours have been plotted for clarity. To aid readability, Kulczynski has been plotted in a log scale. Each distance metric has a distinct geometry, and helps in making \dimmad{} more robust. The correlation metric is omitted from this plot, as it cannot be meaningfully represented within a two-dimensional space.}
    \label{fig:dm}
    \vspace{-5mm}
\end{figure}

A distance metric is a function $d(x, y)$ that quantifies the similarity between two points, obeying three core axioms: identity of indiscernibles, symmetry, and the triangle inequality (refer to Appendix, \cite{chaini_light_2024}). While the Euclidean metric is most commonly used, many other functions satisfy these axioms, each imposing a unique geometry on the feature space (shown in \autoref{fig:dm}). %
While individual distances have been used for AD before (\cite[\eg][]{churova_anomaly_2021, souto_arias_aida_2023}), \dimmad{} uses an ensemble of distance metrics. We chose 16 distance metrics\footnote{Full list of distance metrics is available online: \href{https://github.com/sidchaini/distclassipy/blob/42444f00577030c78c416c7aa3c80689a92e87d9/distclassipy/distances.py\#L99C1-L99C16}{DistClassiPy docs}.} that performed well on light curve classification with \dcpy\  \cite{chaini_light_2024}, %
but we removed two metrics from the set used in \cite{chaini_light_2024} (Additive Symmetric $\chi^2$ \& Maryland Bridge) to increase diversity and decrease redundancy, which is crucial for the ensemble to achieve a robust, consensus-based anomaly score capable of handling different feature space geometries.

\subsection{Multi-metric Anomaly Score}
\dimmad{} is a semi-supervised AD method. It consists of two steps -- the first step (training) involves calculating the centroids of `seen classes' (knowns/inliers) in a given labeled feature set. Then, during test time, it looks at an unseen unlabeled dataset and sorts all of them based on an anomaly score -- giving us a ranking of the most anomalous to least anomalous objects. This semi-supervised approach allows us to incorporate our knowledge of known classes into the anomaly hunt, unlike some other unsupervised AD approaches (for \eg, iForest, LOF, \etc)

Our central hypothesis is that a `true novelty' will be distant from the centroids of all known classes across a majority of distance metrics. Thus, at test time, 
we calculate the distances of all test objects to all known classes for all 16 distance metrics. We then calculate the anomaly score in two aggregation stages:
    (1) For every distance metric, aggregate the distance of the test object to every class centroid by a statistic (class agg: \texttt{min/median}) to get a single-metric score for every metric.
    (2) Then, aggregate these single-metric scores across all distance metrics by another statistic (metric agg: \texttt{median}) to get the final multi-metric score.
Through this, we get a multi-metric anomaly score for all test objects, and rank them from highest (most anomalous) to lowest (least anomalous).

\section{Experiments \& Results} \label{sec:exp}
We test \dimmad{} against a suite of AD algorithms, including the state-of-the-art methods in time-domain astronomy. This set was chosen to represent a variety of approaches common in literature: tree-based methods like the  Isolation Forest  \cite[iForest, ][]{liu_isolation_2008, liu_isolation-based_2012, malanchev_anomaly_2021}, density-based methods like the Local Outlier Factor \cite[LOF, ][]{breunig_lof_2000, lochner_astronomaly_2021}, boundary-based methods like the One-Class Support Vector Machines \cite[OC-SVM, ][]{liu_isolation_2008, liu_isolation-based_2012}, as well as modern deep learning approaches like Autoencoders \cite{rumelhart_parallel_1986, villar_deep-learning_2021}, and the Multiclass Deep support vector data description \cite[MCSVDD, ][]{pmlr-v80-ruff18a, perez-carrasco_multi-class_2023}.

We test our method on the two key discovery tasks described earlier: finding entirely new phenomena (OOD detection) and finding specific rare subtypes (rare in-distribution detection). A key challenge at the start of a survey like LSST is that labeled data will be scarce. Our experimental setup aims to mimic this `data-poor' scenario: we train only on a small number of well-understood object classes. Evaluating performance for the discovery of astrophysical unknowns is non-trivial. Metrics used to assess the power of classifiers are not suited to assess the performance of an AD method. Standard metrics like the AUROC \cite{mcclish_analyzing_1989} evaluate performance across all thresholds, and can be misleading for highly imbalanced datasets. An AUROC score would contaminate the result with the distribution of values at low anomaly scores, which is by definition not an interesting region of the space where to find anomalies. Instead, we evaluate performance using a metric directly tied to the scientific goal: purity \vs{} follow-up budget. Purity (also known as precision) measures the fraction of `true' out-of-class anomalies discovered within the top $N$ ranked candidates, reflecting the reality of limited telescope resources in order to prioritize true novelties for follow-up. Our approach focuses on the crucial high-confidence regime 
relevant for effective astronomical follow-up \cite[\eg ][]{aleo_anomaly_2024}.

For our experiments, we use two datasets of astronomical light curves. First, we use the Extended LSST Astronomical Time-series Classification Challenge \cite[ELAsTiCC, ][]{narayan_extended_2023}, a simulated dataset designed to mimic the six-band ($ugrizy$) data from the upcoming Vera C. Rubin Observatory. Second, we use real data from the Zwicky Transient Facility \cite{bellm_zwicky_2019}, a precursor survey to LSST that provides public data in two optical bands ($gr$). 

We run our models on features derived from the Supernova Parametric Model \cite{villarSupernovaPhotometricClassification2019} fit to light curves with the ALeRCE pipeline software \cite{jainagaAlercebrokerLcClassifier2021}. These features (\eg, amplitude, rise/fall times) are easy and computationally inexpensive to calculate, and crucially, have a low failure rate ($<5\%$), making them robustly available for nearly all objects as compared to the other features which can often have much higher failure rates ($\sim 45\%$), making it a pragmatic choice for real-time anomaly searches with surveys like ZTF and LSST. For the ELAsTiCC data, we use 25 features derived from model fits across the $grizy$ bands, and for the ZTF data, we use 14 features from the $g$ and $r$ bands. Details of this are available on our code repository.

For each run, we first create a balanced training set by downsampling all `known' inlier classes to 
the number of samples in the smallest class. The test set for each run is constructed from unseen inliers and a diverse, stratified sample of `unknown' outlier objects. While not astrophysically representative, the test set's high anomaly fraction ($\sim47\%$, spread equally over many anomalous unknown classes) is a deliberate choice to create a robust benchmark, ensuring a statistically stable comparison of each algorithm's ranking performance. However, we also tested on a test dataset composed of $>98\%$ inliers to confirm the robustness of the results we present here (described in \autoref{app:newtest}). All models are evaluated on the exact same train/test splits. We perform Monte Carlo cross-validation \cite{xu_monte_2001}, and run each experiment 20 times with different random seeds to ensure statistical robustness.

\subsection{Out-of-Distribution (Inter-Class) Anomaly Detection} \label{subsec:ood}

We first test \dimmad{} at finding `unknown unknowns'. We simulate this by training on well-understood classes of objects (knowns) and treating physically distinct classes as unseen, OOD anomalies (unknowns).

\begin{figure}[htb!] %
    \centering
    \includegraphics[height=0.22\textheight]{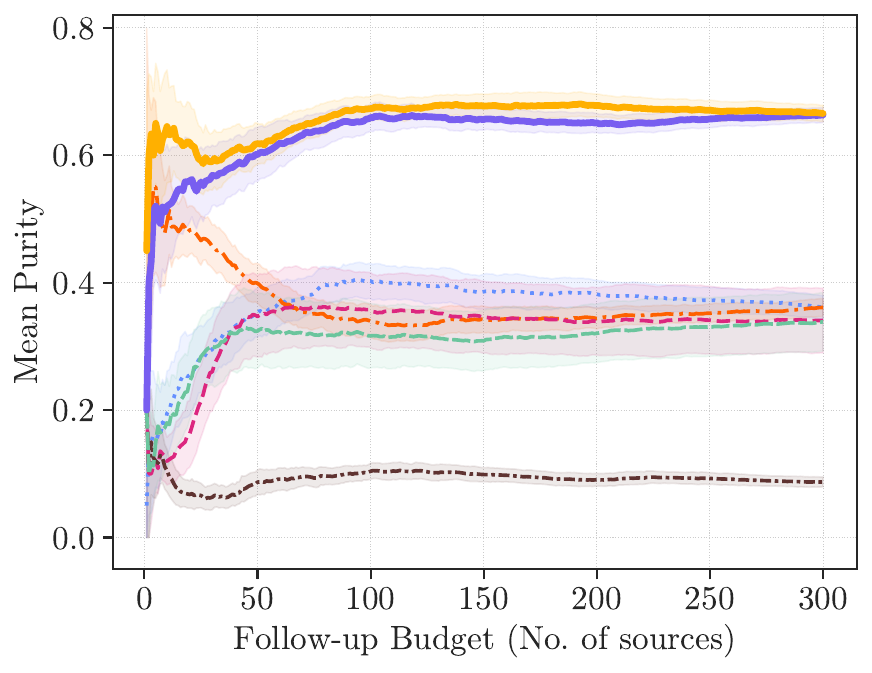}
    \includegraphics[height=0.22\textheight]{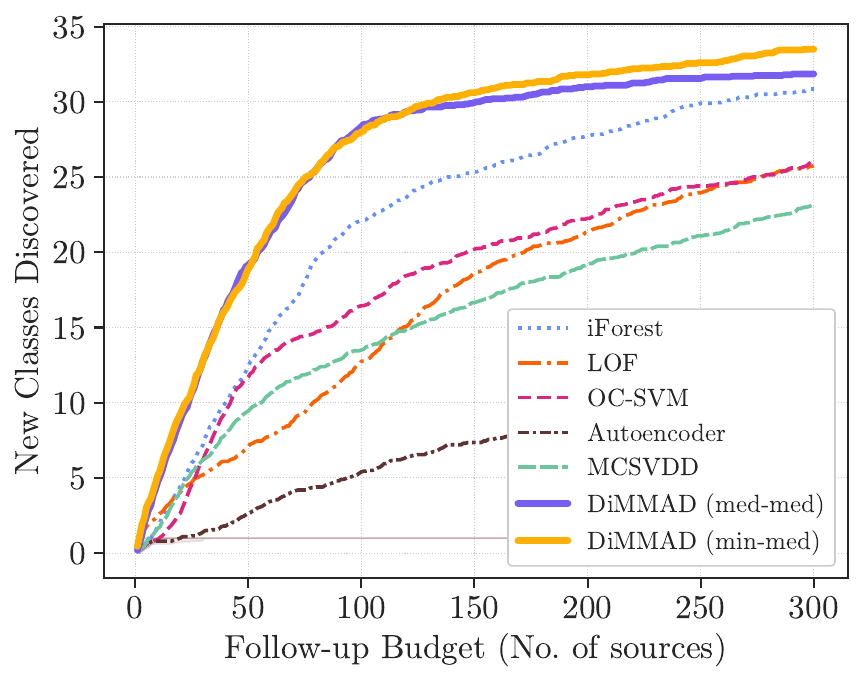}
    \caption{Anomaly Detection Performance of our method (\dimmad), compared to other standard AD methods. 
    The shaded region denotes $1\sigma$ errors from Monte Carlo cross-validation. %
    \emph{Left}: the mean purity of OOD anomalies within the top $N$ candidates on the ELAsTiCC data. The \dimmad{} methods (solid lines) maintain significantly higher purity. \emph{Right}: the cumulative number of new, unique OOD classes discovered on the ELAsTiCC data. \dimmad{} discovers a greater, more diverse set of anomalies more efficiently.}
    \label{fig:exp1}
\end{figure}

\paragraph{Simulated LSST Data (ELAsTiCC)}
To create a controlled environment, we train the models on four classes of variable stars (cepheids, RR lyrae, delta scutis and eclipsing binaries) and task them with identifying 34 unseen transient classes (supernovae, TDEs, \etc) as anomalies. The results for this are shown in \autoref{fig:exp1} for two variants of \dimmad{} (cluster aggregators (\texttt{min/median}) with \texttt{median} as the metric agg. The purity plot (\autoref{fig:exp1} left panel%
) shows that both \dimmad{} methods are the best performing in the lower budget regime, and consistently maintain a purity above 60\% within a realistic follow-up budget of the first few hundred objects. Furthermore, \autoref{fig:exp1} right panel %
shows that \dimmad{} identifies a diverse range of new phenomena more rapidly than any other method, maximizing the potential for broad scientific discovery.

\begin{figure}[htb!] %
    \centering
    \hspace{5em}
    \includegraphics[height=0.22\textheight]{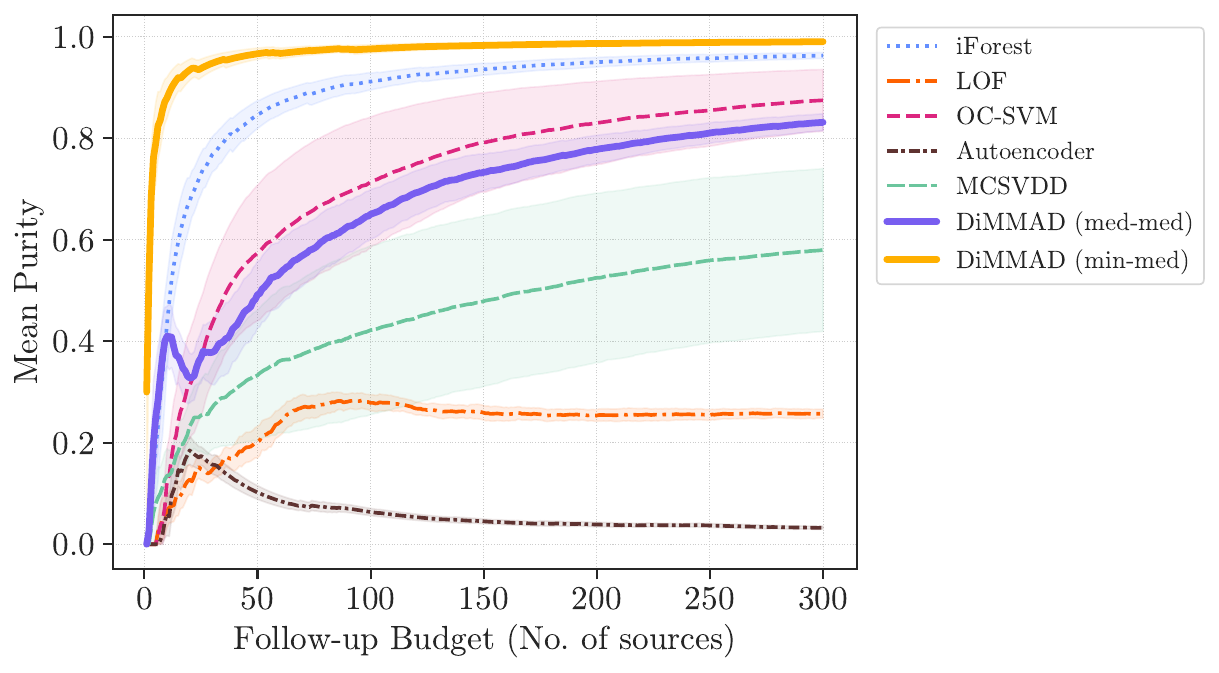}
    \caption{
    As \autoref{fig:exp1} (\emph{left}) but for the case of real ZTF (ALeRCE) data.}
    \vspace{-4mm}
    \label{fig:exp1c}
\end{figure}

\paragraph{Real ZTF Data (ALeRCE)}
We validate these findings on real ZTF data, from the light curve feature set calculated by ALeRCE \cite{perez-carrasco_alert_2023}. We train on 11 common variable and stochastic classes and treat four supernova classes as the unseen OOD sample. We see that one variant of \dimmad{} (\texttt{min} cluster agg, \texttt{median} metric agg) is again the best performing. The other \dimmad{} variant (\texttt{med-med}) performs comparably to other methods like iForest and OC-SVM (\autoref{fig:exp1c}).

\subsection{Rare In-Distribution (Intra-Class) Anomaly Detection}
We also tested a more challenging scenario: finding rare subtypes within a broader family of objects. For this task, we trained on 16 known supernova subtypes and defined 3 rare pair-instability supernova models as the `unknown' class. While \dimmad{} remains competitive, we find that iForest often shows superior performance. This suggests that while our multi-metric consensus excels at finding OOD objects, other approaches may be better suited for finding rare examples within a known population. Our hypothesis for this difference is as follows - the iForest finds isolated points located in the sparse regions of the manifold, while DiMMAD's distance based approach explicitly finds objects that are robustly distant from all known centroids across multiple geometries, making it exceptionally effective for off-manifold (OOD) detection but less sensitive to subtle, on-manifold variations. More details for this experiment are provided in \autoref{app:exp5}.

\section{Conclusion}
We introduced \dimmad{}, a novel, semi-supervised anomaly detection method that leverages an ensemble of distance metrics to identify novel astronomical phenomena. Our work demonstrates that \dimmad{} is a powerful tool for Out-of-Distribution detection, a crucial task for maximizing serendipitous discoveries in large surveys like LSST. The strength of our approach lies in its demand for a geometric consensus. While standard algorithms may flag extreme in-sample objects located in sparse regions of the feature space, \dimmad{} requires a candidate to be robustly distant from all known classes from the perspective of multiple, diverse geometries. This makes it particularly effective at identifying truly novel, off-manifold objects. A key advantage of our semi-supervised approach is its ability to dynamically improve. Once a new class of anomaly is discovered and characterized through follow-up, its centroid can be added to the set of `knowns'.

For the Rubin LSST community, \dimmad{} offers an interpretable and computationally efficient method to filter through vast alert streams, enriching the top of ranked candidate lists with true novelties and optimizing follow-up resources. Our implementation is open-source, and available within the \dcpy{} package, \footnote{\href{https://github.com/sidchaini/distclassipy/}{https://github.com/sidchaini/distclassipy/}} along with all code to reproduce the results of this paper.\footnote{\href{https://github.com/sidchaini/dimmad/}{https://github.com/sidchaini/dimmad/}}

\newpage

\begin{ack}
SC acknowledges support received from the University of Delaware Doctoral Fellowship of Excellence, and the NASA FINESST program, Grant 80NSSC25K0312. FBB is supported in part by NSF AST Award Number 2511639. AM is supported in part by NASA Grant 80NSSC24M0020.

\end{ack}

\renewcommand*{\bibfont}{\small}
\bibliographystyle{aasjournal}
\bibliography{references, dcpyrefs, references-2}

\appendix
\section{Performance on Individual Out-of-Distribution Classes} \label{app:newtest}

To understand how \dimmad{} performs on specific types of novelties, we repeated the OOD experiment from \autoref{subsec:ood} on ELAsTiCC with the same `known' set of 4 variable stars (CEP, RR, DSCT, EB), but restricted the test set to contain only one `unknown' transient class at a time, comprising <2\% of the total test set. This tests the ability of each algorithm to recover a single, rare type of OOD signal. 

\begin{figure}[htp!] %
    \centering
    \includegraphics[height=0.22\textheight]{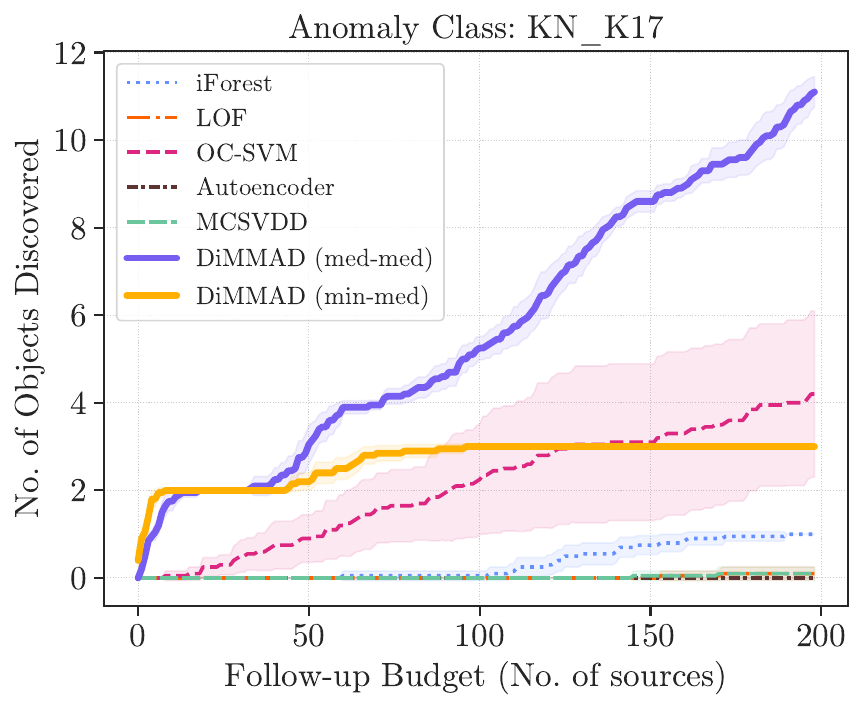}
    \includegraphics[height=0.22\textheight]{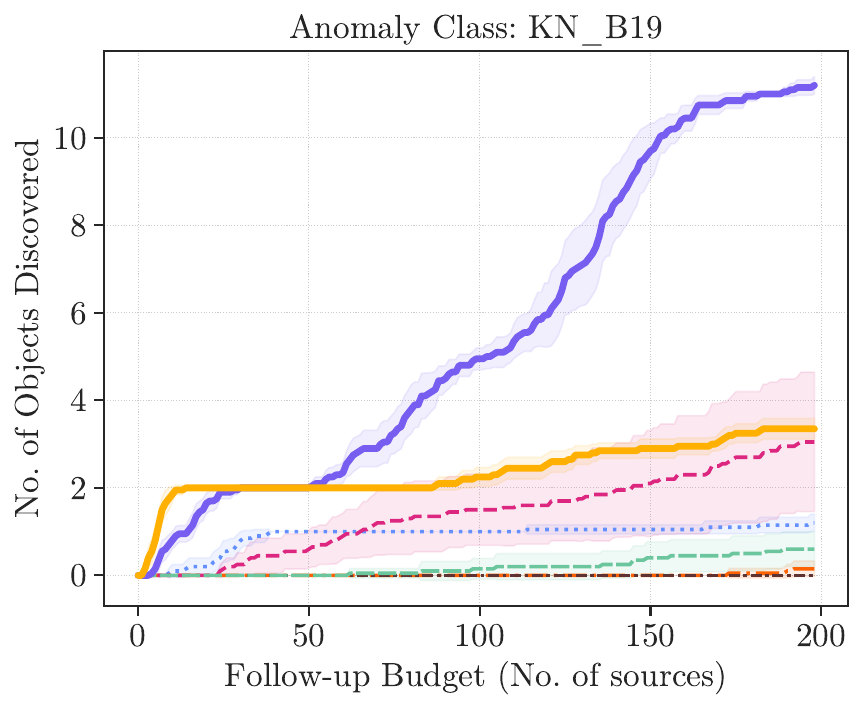}
    \includegraphics[height=0.22\textheight]{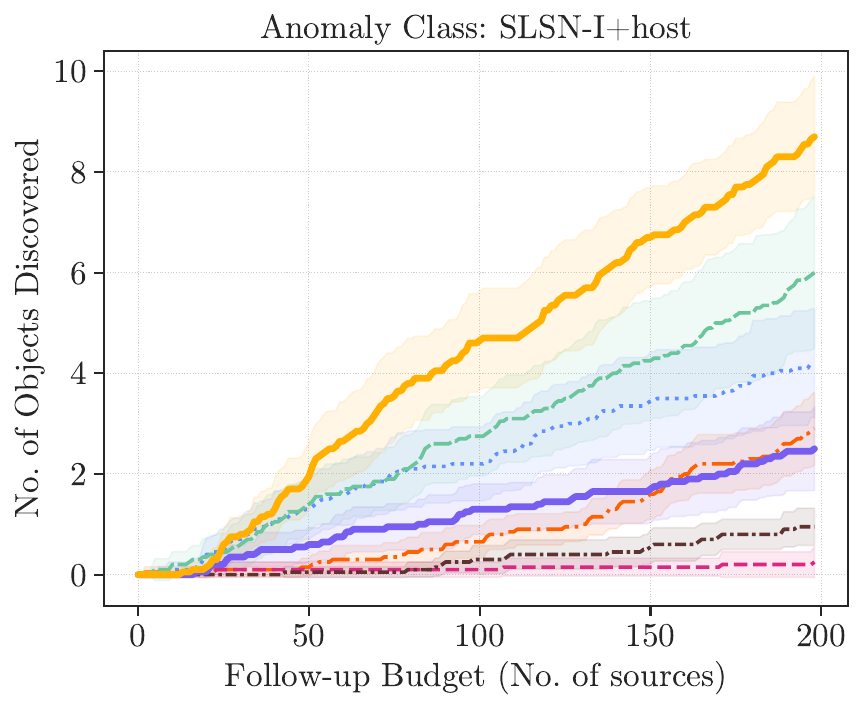}
    \includegraphics[height=0.22\textheight]{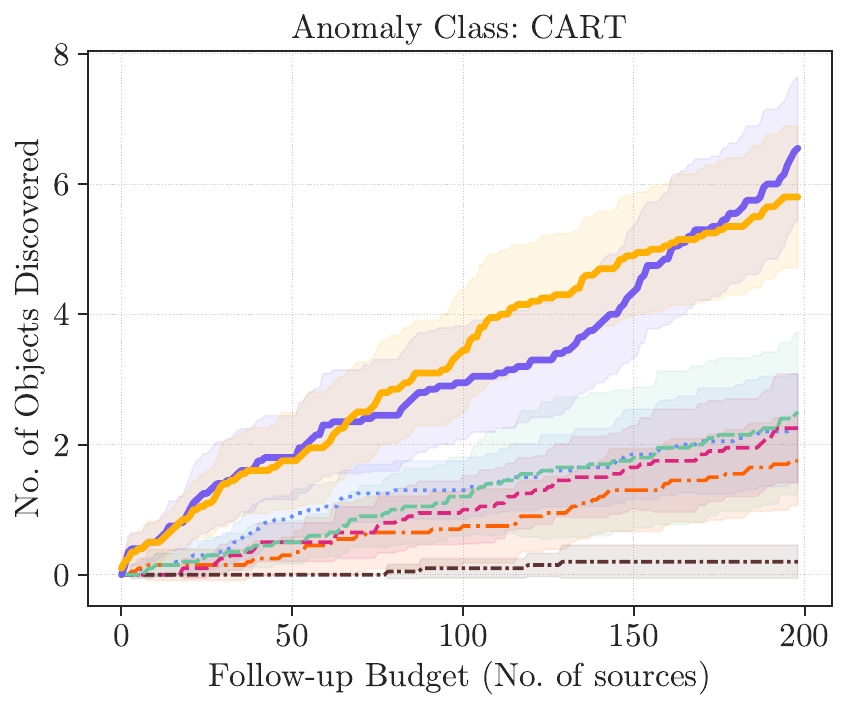}

    \includegraphics[height=0.22\textheight]{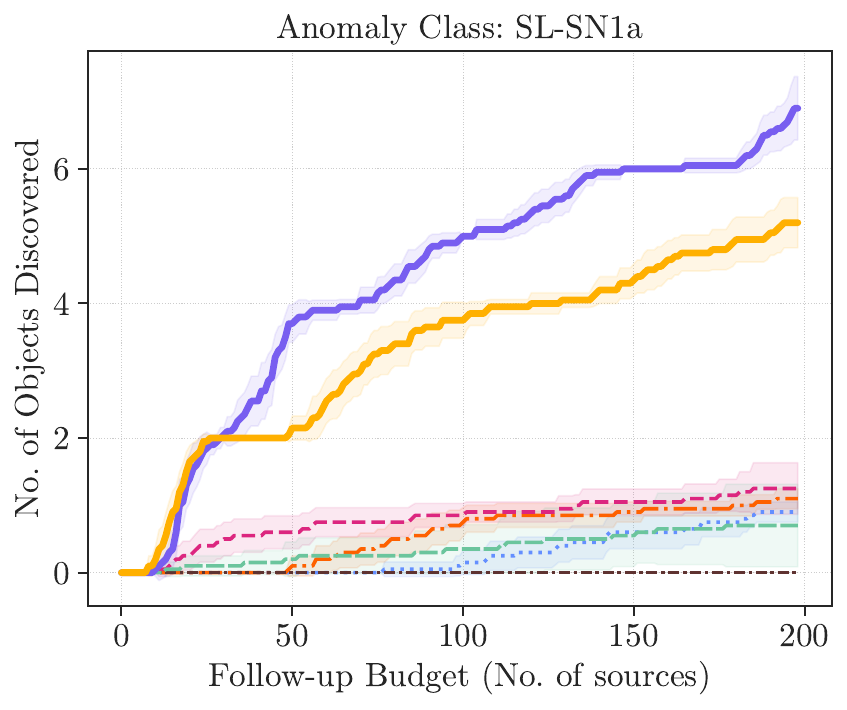}
    \includegraphics[height=0.22\textheight]{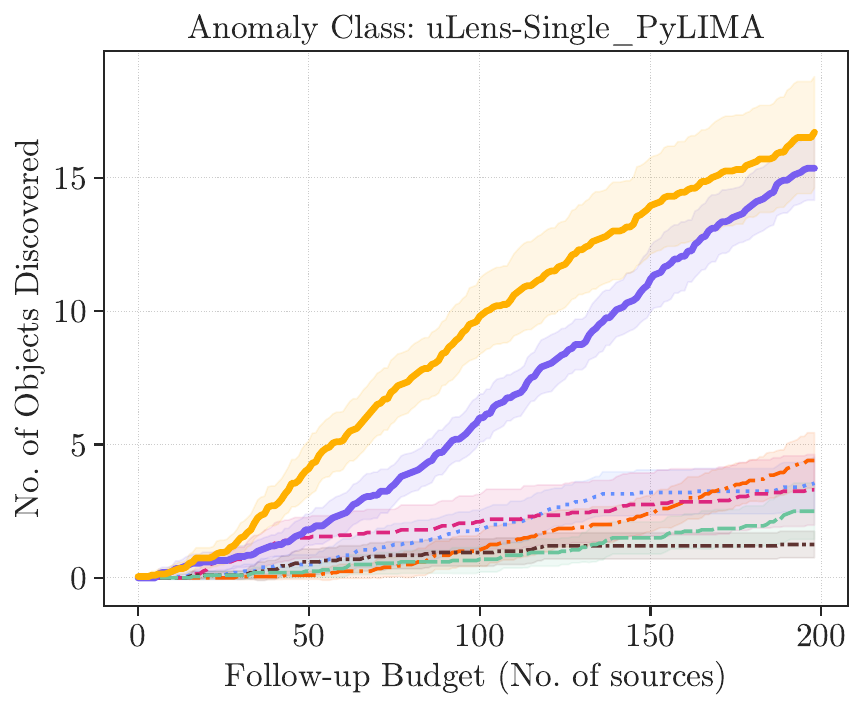}
    \includegraphics[height=0.22\textheight]{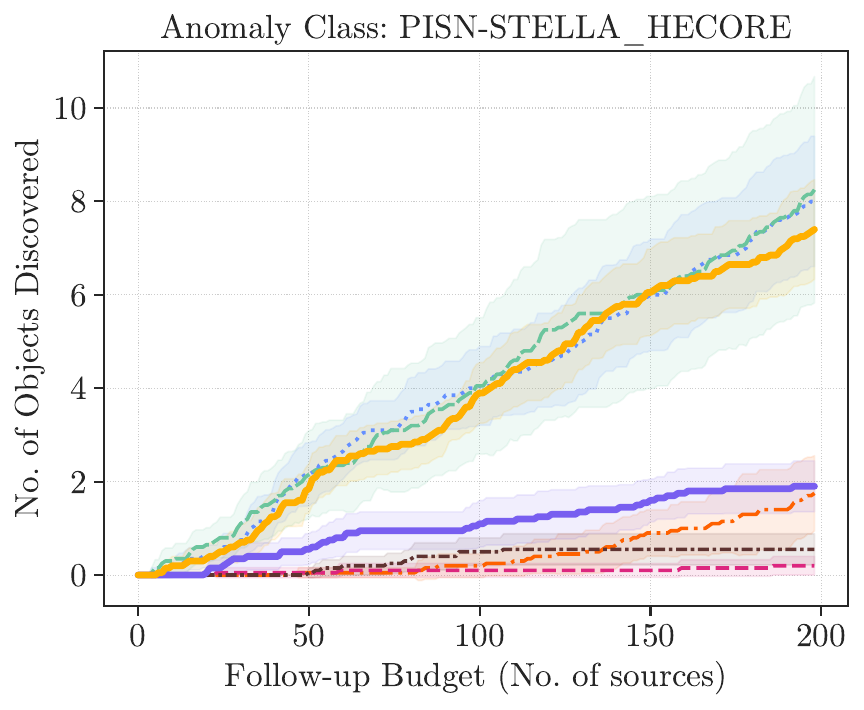}
    \includegraphics[height=0.22\textheight]{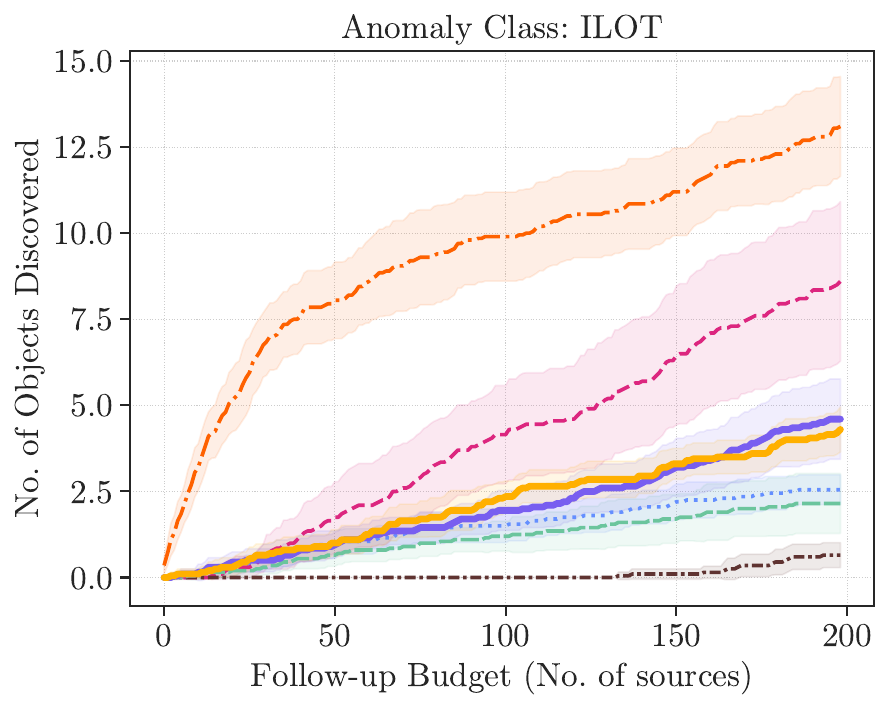}

    \caption{
    Cumulative number of objects discovered vs. follow-up budget for individual OOD classes from the ELAsTiCC dataset. Each panel represents a separate experiment where the titled class was the sole anomaly type. \dimmad{} (bold orange/purple) consistently performs among the best methods for most classes of OOD anomalies across a diverse range of physical phenomena
    }
    \label{fig:exp6}
\end{figure}

\autoref{fig:exp6} shows the mean cumulative number of objects discovered for a selection of physically diverse and rare anomalous classes over 20 runs of Monte Carlo cross validation. The discovery rate is much lower for all models, but the \dimmad{} methods consistently rank among the top performers, demonstrating particular strength in identifying rare explosive transients like Kilonovae (KN), Superluminous Supernovae (SLSN) and Calcium Rich Transients (CART) as well as other rare events like Strongly Lensed Type Ia Supernovae (SL-SN1a) and Microlensing events (uLens). For some classes, other methods like MCSVDD, iForest and LOF perform as well (Pair-Instability Supernovae; PISN) or better (Intermediate Luminous Optical Transients; ILOT), highlighting that no single algorithm is universally optimal. However, across the full spectrum of anomalies, the \dimmad{} ensemble approach provides a consistently high rate of discovery, reinforcing its utility as an efficient, interpretable general-purpose method for finding the unknown.

\section{\dimmad{} Performance on Rare In-Distribution Anomalies} \label{app:exp5}

\begin{figure}[H] %
    \centering
    \vspace{2em}
    \includegraphics[height=0.22\textheight]{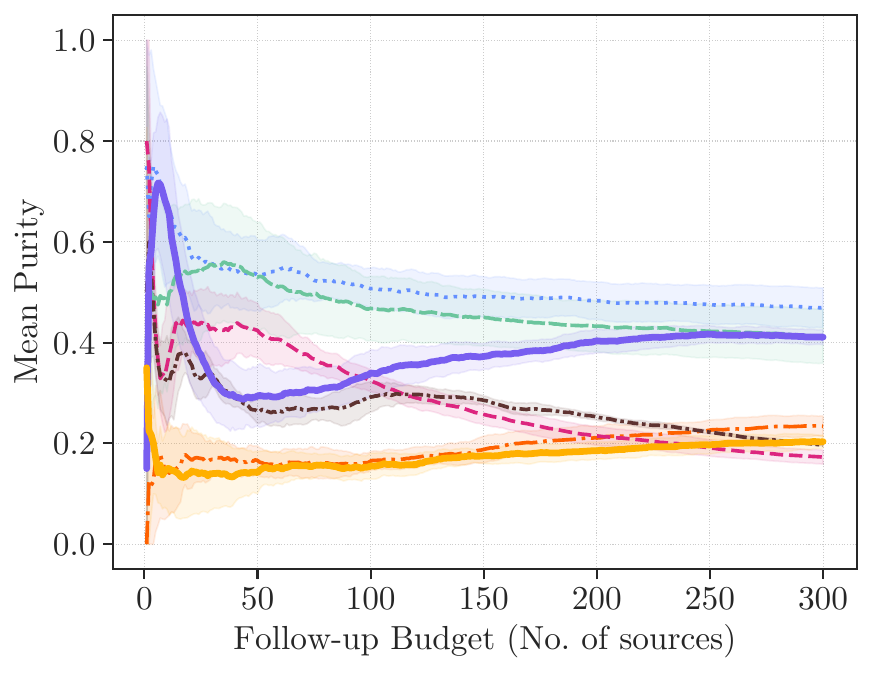} 
    \hspace{1em}
    \includegraphics[height=0.22\textheight]{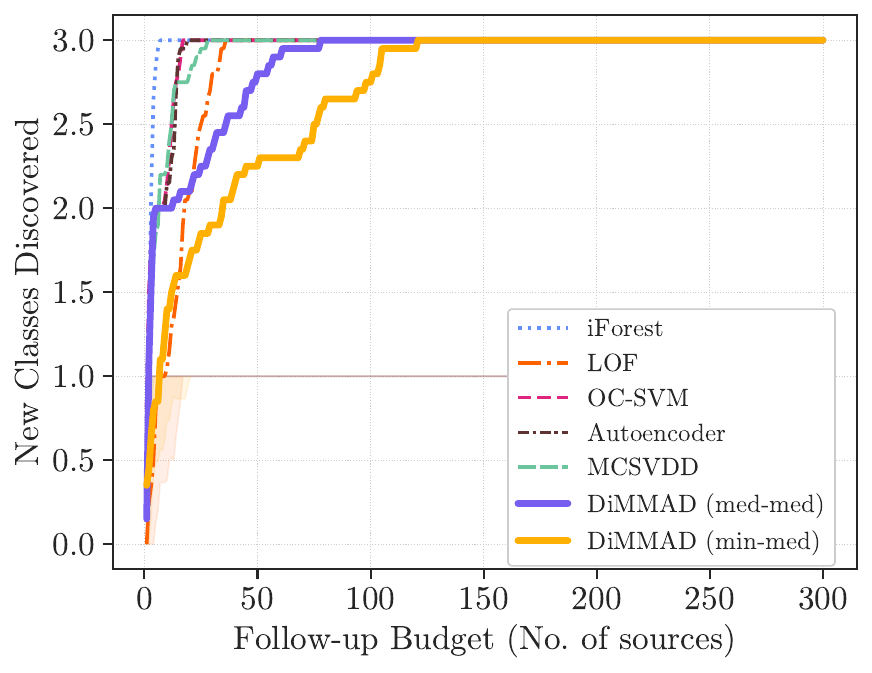}
    \caption{Anomaly Detection Performance of our method (\dimmad), compared to other standard AD methods on the ELAsTiCC data as in \autoref{fig:exp1}, but for Rare In-Distribution Anomalies.
    The shaded region denotes the $1\sigma$ errors from Monte Carlo cross-validation. 
    }
    \label{fig:exp5}
\end{figure}

To test \dimmad{}'s performance on subtle, intra-class anomalies in ELAsTiCC, we trained our models on 16 known classes of supernova models (for Supernova Types Ia, Iax, Ib, Ic, II, IIb, IIn, and SLSN-I), and tried finding 3 predicted classes of rare pair-instability supernova models (PISN) as the `unknown' class. The results (\autoref{fig:exp5}) show that while \dimmad{} is competitive, iForest provides a higher and more stable purity for this on-manifold detection task. This highlights that the optimal anomaly detection method is task-dependent. While our multi-metric ensemble excels at finding `unknown unknowns' (OOD), other approaches may be better suited for finding the `known unknowns' (rare in-distribution).

\end{document}